\documentstyle[12pt]{article}

\begin{document}
\renewcommand{\thefootnote}{\fnsymbol{footnote}}
\begin{titlepage}
\begin{flushright}
{\sc Report\ $\sharp$\ \ \ RU95-1-B}\\
\end{flushright}
\vspace{.6cm}
\begin{center}
{\Large{\bf LARGE CONFORMAL DEFORMATIONS}}\\[5pt]
{\Large{\bf AND SPACETIME STRACTURE}}\\[30pt]
{\sc Gregory Pelts
\footnote{Supported by the DOE grant {\em DOE-91ER4651 Task
B}}}\\[2pt]
{\it
Department of Physics\\
The Rockefeller University\\
1230 York Avenue\\
New York, NY 10021-6399}\\[40pt]
{\sc Abstract}\\[12pt]
\parbox{13cm}{
We demonstrate in detail how the space of two-dimensional
quantum field theories can be parametrized by
off-shell closed string states.
The dynamic equation corresponding to the condition of
conformal invariance includes an infinite number of higher order
terms,
and we give an explicit procedure for their  calculation.
The symmetries corresponding to
equivalence relations of CFT are described.
In this framework
we show how to perform nonperturbative analysis in
the low-energy limit and prove that it corresponds to the
Brans-Dicke theory of gravity interacting with a skew symmetric
tensor
field.
\\[2ex]
{\em Talk presented at the G\"ursey memorial conference, Istanbul,
Turkey, June, 1994. }
}
\end{center}
\end{titlepage}
\section{Introduction}
Classical closed string states are believed to be associated with
quantum conformal field theories in two dimensions (CFT),
which are usually defined as theories of the single string
moving  in some nontrivial spacetime background.
The condition of anomaly cancellation leads to the
so-called $\beta$-function equation on the background fields.
The main advantage of this approach is its more or less explicit
connection to spacetime geometry, and the main drawback is that it
usually
focuses only on massless fields.
Treatment of massive fields is problematic
and, therefore, characterization of dynamical degrees of freedom is
obscure.
Symmetries also are not explicit because
classically equivalent CFT may correspond to inequivalent
quantum theories.
Approaches~[1-5] based on the
operator formalism~[6] encounter
problems dealing with ambiguity of
the vertex operator commutator  due to contact singularity of
their $T$-product.
As we will see, this ambiguity is principal as
it actually makes the string theory nonlinear.

\section{Vertex operators}
We will consider CFT as a family of amplitudes
$\langle 0 \rangle_{\Sigma}$ assigned to Riemann surfaces
$\Sigma$ and obeying sewing property
(see details in~[7]).
It is similar to the Segal's definition~[8].
In this formalism vertex operators $\Psi(z_0)$
inserted at the point $z_0$ can be defined as a family of states
$\langle\Psi(z_0)\rangle_D\in{\cal H}^{\partial D}$
associated with disc like environments $D$ of $z_0$
obeying the condition
$$
{\langle\Psi(z_0)\rangle}_{D_2}=
Sp_{\partial D_1}{\langle 0 \rangle}_{D_2\setminus D_1}\otimes
\langle\Psi(z_0)\rangle_{D_1}
\hspace{1em}(D_1\subset D_2).
 $$
Here $Sp_{\Gamma}$ denotes contraction of amplitudes corresponding
to sewing of Riemann surfaces along the contour $\Gamma$.
The $T$-product of such vertex operators can be defines as
\begin{equation}
\langle\Psi_0(z_0)\cdots\Psi_n(z_n)\rangle_{\Sigma}
=Sp_{\partial\Sigma_{\rm ext}}
\bigotimes_{i=0}^n
{\langle\Psi_i(z_i)\rangle}_{D_i}\otimes{\langle 0
\rangle}_{\Sigma_{\rm
ext}}.
\label{t-p}
\end{equation}
Here $D_i$ are nonintersecting environments of $z_i$ and
$\Sigma_{\rm ext}$ is
their complement in $\Sigma$.

The Virasoro algebra does not have a bounded natural representation
in  ${\cal H}^{\partial D}$. The conformal transformations
deform the boundary of the disk and, therefore,
corresponding to them linear operators are not automorphisms.
However, we can define such a representation
in the space of vertex operators, which is independent of the
position
of the boundary.

\section{Infinitesimal Deformation of CFT}
Infinitesimal deformations of amplitudes can be parametrized
by (1,1)-primary fields
$$
{\langle 0 \rangle}_{\Sigma}\ \longrightarrow \ {\langle 0
\rangle}_{\Sigma} +\delta{\langle 0 \rangle}_{\Sigma},\hspace{1em}
\delta{\langle 0 \rangle}_{\Sigma}=
\frac{1}{\pi}\int_{\Sigma}
{\langle\Psi(z)\rangle}_{\Sigma}
\,d^2\!z.
 $$
Formally we can parametrize vertex
operators of the deformed theory by vertex operators of the initial
theory
as
$$
\langle\Upsilon\rangle_{D}
\ \longrightarrow \
\langle\Upsilon\rangle_{D}
+ \delta\!\langle\Upsilon\rangle_{D}
,\
\delta\!
\langle\Upsilon(z_0)\rangle_{D}=
\frac{1}{\pi}\int_{D}
\langle\Psi(z)\Upsilon(z_0)\rangle_{D}
\,d^2\!z .
 $$
However, in general,
the integral here
may be divergent because of the contact singularity of the
$T$-product.
The simple cutoff regularization
$$
\delta_R\langle\Upsilon(z_0)\rangle_{\Sigma}=
\frac{1}{\pi}\int_{\Sigma\setminus D_{z_0,R}}
\langle\Psi(z)\Upsilon(z_0)
\rangle_{\Sigma}
\,d^2\!z
,\
D_{z_0,R}=\left\{z\in\Sigma,\, |z-z_0|\leq R\right\}
 $$
will violate sewing properties for small disks.
Instead we will make an overage of such regularizations over
infinitely
small cutoff radiuses
\begin{equation}
\delta\!\langle
\Upsilon(z_0)
\rangle_{\Sigma}=
\int_0^{\infty}\delta_r
\langle\Upsilon(z_0)
\rangle_{\Sigma}d\!\mu(r).
\label{reg}\end{equation}
Here $d\!\mu$ is a generalized measure in ${\bf R}_+$ having support
in $0$ and integrable in a  product with all the functions having a
finite degree singularity at $r=0$.
Such a measure exists  and is fully described by the function
$$
\Lambda(\alpha)=\int_0^{\infty} r^{2\alpha} \mu (r)\,d\!r
\hspace{1em}
(\alpha\in{\bf R}) ,
 $$
satisfying
$$
\Lambda(0)=1,\hspace{1em}\Lambda(\alpha)=0\hspace{1em}
(\alpha>A)
 $$
for some positive $A$.

\section{Residue-like operations}
We will call local linear operators  from
the space of functions on
$\Sigma^{N+1}$ having diagonal singularities
to the space of functions on $\Sigma$ residue-like operators
of rank N and denote them as
$$
G(z_0)={\cal R}_{z_N=\cdots=z_0}F(z_1,\ldots,z_0)\
\mbox{ or }\
G(z_0)={\cal R}_{\bar{z}_N=\cdots=\bar{z}_0}F(z_1,\ldots,z_0).
 $$
Using $T$-product~(\ref{t-p}) we can define representation of
residue-like operations by
multilinear products in the space  of vertex operator functions:
\begin{eqnarray}
\{\Upsilon_i\}_{i=1}^N\ \longrightarrow \ \
\Upsilon&=&{\cal
R}_{z_N=\cdots=z_0}\Upsilon_0\ldots\Upsilon_N,\nonumber\\
\langle\Upsilon(z_0)\rangle_{\Sigma}
&\stackrel{\rm def}{=}&
{\cal R}_{z_N=\cdots=z_0}
\langle\Upsilon_0\ldots\Upsilon_N\rangle_{\Sigma}.
\nonumber\end{eqnarray}
The deformation~(\ref{reg})
induce the deformation of this representation
\begin{equation}
\delta\!{\cal R}_{z_n=\cdots=z_0}
\Upsilon_0(z_0)\cdots\Upsilon_n(z_n)
={\cal R}_{z\doteq z_n=\cdots=z_0}
\Upsilon_0(z_0)\cdots\Upsilon_n(z_n)\Psi(z).
\label{rd}\end{equation}
Here ${\cal R}_{z\doteq z_n=\cdots=z_0}$ is
a next rank residue-like operation
defined as
\begin{equation}
{\cal R}_{z\doteq z_n=\cdots=z_0}F={\cal R}_{z_n=\cdots=z_0}
\frac{1}{\pi}\int_{\Sigma}F\,d^2\!z
-\frac{1}{\pi}\int_{\Sigma}
{\cal R}_{z_n=\cdots=z_0}F
\,d^2\!z  .
\label{sc}\end{equation}
It is, indeed, a residue-like operation, because  the right part
of~(\ref{sc})
does not depend on the area of integration as far as it includes
$z_0$.
We will call this operation a successor of ${\cal
R}_{z_{z_n=\cdots=z_0}}$.
A successor of antiholomorphic derivative can be shown to be
$$
\partial_{\bar{z}\doteq\bar{z}_0}F
=-{\rm Res}_{z=z_0}F.
 $$
Here ${\rm Res}_{z= z_i}$ is a generalized residue operation defined
for nonholomorphic functions as
$$
{\rm Res}_{z= z_0}\frac{(z-z_0)^k}{|z-z_0|^{2\alpha}}=
\frac{\Lambda(k-\alpha)}{k!(k+1)!}\delta_{k,-1}
\hspace{1em}
(\alpha\in{\bf R}, k\in{\bf Z}) .
 $$

\section{Finite Deformations}
Let deformed amplitudes be defined by the formula
$$
{\langle 0 \rangle}_{\Sigma}^{\Psi} =
\left\langle
\exp\frac{1}{\pi}\int_{\Sigma}\Psi
\,d^2\!z
\right\rangle_{\Sigma}.
 $$
Here $\Psi$ is some vertex operator function
(not necessarily primary),
and the contact divergences are regularized by the
method~(\ref{reg}).
Then the sewing property is automatically
satisfied, and only the condition of conformal invariance
remains to be implemented.
Hereafter we mark all the deformed objects
with the superscript symbol of the vertex operator function
parameterizing the deformation.
For parametrization of deformed vertex operators
we will use the formula
$$
\langle \Upsilon(z_0)\rangle_{\Sigma}^{\Psi}=
\left\langle
\Upsilon(z_0)\exp
\frac{1}{\pi}\int_{\Sigma}\Psi(z)
\,d^2\!z
\right\rangle_{\Sigma} .
 $$
The energy-momentum tensors  for the
family of deformed theories parametrize
by scaled vertex operator function $\tau\Psi$ $(\tau\in{\bf R})$
can be shown to obey the differential equation
\begin{eqnarray}
\frac{\partial}{\partial\tau}T_{zz}^{\tau\Psi}(z)
&=&
{\cal B}_{z_1=z}^{\tau\Psi}\Psi(z_1)T_{zz}^{\tau\Psi}(z)+
{\cal A}_{\bar{z}_1=\bar{z}}^{\tau\Psi}
\Psi(z_1)T_{z\bar{z}}^{\tau\Psi}(z)\nonumber\\
\frac{\partial}{\partial\tau}T_{z\bar{z}}^{\tau\Psi}(z)
&=&
{\cal A}_{z_1=z}^{\tau\Psi}\Psi(z_1)T_{zz}^{\tau\Psi}(z)
+{\cal B}_{\bar{z}_1=\bar{z}}^{\tau\Psi}\Psi(z_1)
T_{z\bar{z}}^{\tau\Psi}(z)+\Psi(z) .
\label{cf}\end{eqnarray}
Here ${\cal A}_{z_1=z}$, ${\cal B}_{z_1=z}$ are residue-like
operations satisfying
$$
\partial_{z}{\cal A}_{z_1=z}+\partial_{\bar{z}}{\cal B}_{z_1=z}=
{\rm Res}_{z_1=z}+{\rm Res}_{z=z_1}.
 $$
Differentiating~(\ref{cf}) with respect to $\tau$
and applying~(\ref{rd})
we can recurrently calculate all the higher derivatives
of the energy-momentum tensor
and then substitute them to  the Taylor expansion.
Thus we have a perturbative formula for $T^{\Psi}$ with higher order
terms expressed through residue-like operations ${\rm Res}_{z=z_0}$,
${\cal A}_{z=z_0}$,  ${\cal B}_{z=z_0}$  and their
successors~(\ref{sc}).
Details on calculation of these operations can be found in~[7].
The transformation of energy-momentum tensor
\begin{eqnarray}
T_{zz}^{\Psi,\Phi}=T_{zz}^{\Psi}-
\partial_{z}^{\Psi}\Phi_z,
\hspace{1em}&&
T_{z\bar{z}}^{\Psi,\Phi}=
T_{z\bar{z}}^{\Psi}+\partial_{\bar{z}}^{\Psi}\Phi_z,
\nonumber\\
T_{\bar{z}\bar{z}}^{\Psi,\Phi}=
T_{\bar{z}\bar{z}}^{\Psi}-
\partial_{\bar{z}}^{\Psi}\Phi_{\bar{z}},
\hspace{1em}&&
T_{\bar{z} z}^{\Psi,\Phi}=T_{\bar{z} z}^{\Psi}
+\partial_{z}^{\Psi}\Phi_{\bar{z}}
\nonumber\end{eqnarray}
does not affect translation operators $L_{-1}$, $\overline{L}_{-1}$.
If the theory is conformally symmetrical, there exists $\Phi_z$,
$\Phi_{\bar{z}}$ trivializing the
contradiagonal components of $T^{\Psi,\Phi}$.
Then the diagonal components  of $T^{\Psi,\Phi}$,
will be (anti)holomorphic. They can be used for calculation
of the deformed Virasoro representation.

\section{Symmetries}
The transformations of $\Psi$
\begin{equation}
\delta_{\xi}\Psi(z)=\partial_{\bar{z}}\xi_z(z)+
{\rm Res}_{z_1=z}\xi_z(
\hspace{0.5em}\underbrace{\hspace{-0.5em}
z_1)\Psi(z
\hspace{-0.5em}}_{asym} \hspace{0.5em}
)+ z\leftrightarrow\bar{z} + O(\Psi^2)\label{sym}
\end{equation}
can be shown not to affect equivalence classes of theories.
They are paramet\-rized by the pair
of vertex operator functions
$\xi=(\xi_z,\xi_{\bar{z}})$.
The corresponding {\em covariant\/} transformation
of vertex operators are
$$
\hat{\xi}\Upsilon(z_0)={\rm Res}_{\bar{z}=\bar{z}_0}
\xi_z(z)\Upsilon(z_0)+
\frac{1}{2}{\rm Res}_{z_2\doteq
z_1=z_0}\Psi(z_2)\xi(z_1)\Upsilon(z_0)
+z\leftrightarrow\bar{z} + O(\Psi^2)  .
 $$
The symmetry transformation of $\Phi$ is defined by the requirement
for the energy-momentum tensor $T^{\Psi,\Phi}$ to transform
covariantly.
Note that the commutator of such symmetries depend on $\Psi$
and regularization parameters. Some symmetries related to global
spacetime transformations where also described in [9,10]

\section{Linear approximation}
In the linear approximations the
energy-momentum tensor and symmetries can be shown to be
\begin{eqnarray}
&&T_{z\bar{z}}^{\Psi,\Phi}={\cal O}\Psi
-\overline{L}_{-1}\Phi_z,\hspace{1em}
T_{\bar{z}z}^{\Psi,\Phi}
=\overline{\cal O}\Psi-L_{-1}\Phi_z,
\nonumber\\&&
T_{zz}^{\Psi,\Phi}=T_{zz}+L_{-1}\Phi_z,\hspace{1em}
T_{\bar{z}\bar{z}}^{\Psi,\Phi}=
T_{\bar{z}\bar{z}}+\overline{L}_{-1}\Phi_{\bar{z}},
\nonumber\\&&
\delta_{\xi}\Psi=-\bar{L}_1\xi_z-L_1\xi_{\bar{z}}, \hspace{1em}
\delta_{\xi}\Phi_z={\cal O}\xi_z,\hspace{1em}
\delta_{\xi}\Phi_{\bar z}=\overline{\cal O}\xi_{\bar{z}} .
\nonumber\end{eqnarray}
Here
$$
{\cal O}={\bf 1}+
\sum_{j=0}^{\infty}\frac{
(L_{-1})^j L_{j}}{(j+1)!},\hspace{1em}
\overline{{\cal O}}={\bf 1}
+\sum_{j=0}^{\infty}\frac{(\bar{L}_{-1})^j
\bar{L}_{j}}{(j+1)!}.
 $$
Then the equations
$$
T_{z\bar{z}}^{\Psi,\Phi}=T_{\bar{z}z}^{\Psi,\Phi}=0
 $$
are satisfied if $\Phi$ is trivial and
$\Psi$ is a primary field.
It corresponds to the deformations of the Virasoro
operators
$$
L_k^{\Psi}=L_k + \frac{1}{2\pi i}\oint_{\Gamma}\Psi(z-z_0)^k d\!{\bar
z},
\hspace{1em}
\bar{L}_{k}^{\Psi}=
\bar{L}_k
+\frac{1}{2\pi i}\oint_{\Gamma}
\Psi (\bar{z}-\bar{z}_0)^k d\!z,
 $$
which is equivalent to the deformations proposed in~[1].
Some deformations corresponding to nonprimary fields were first found
in~[11,12] in the low-energy limit.
This relaxation of the equations of motion
is compensated by the symmetries
and does not create additional physical degrees of freedom.

\section{Low-energy limit}
Let us consider deformations corresponding to the vertex operator
function
$
\Psi=H_{\nu\mu}\partial X^{\nu}\bar{\partial} X^{\mu}(z^{\prime})
$,
where $H_{\nu\mu}$ is a Hermitian matrix with slowly-varying
coefficients.
Then the coordinate vertex operators can be shown to have the obey
\begin{eqnarray}
\langle
X^{\nu}(z^{\prime})X^{\mu}(z)
\rangle_{\Sigma}^{\Psi}
&\approx&-
2\langle
:\!g^{\nu\mu}(z)\! :
\rangle_{\Sigma}^{\Psi}
\ln|z^{\prime}-z|,
\nonumber\\
\bar{\partial}^{\Psi}\!\partial^{\Psi}\!:\!\psi\! :
&\approx&
:\!\left(
\psi_{;\nu\mu}+i\psi^{;\eta}d\omega_{\eta\nu\mu}
\right)
\partial^\Psi\! X^{\nu}\bar{\partial}^\Psi\! X^{\mu}\!: .
\nonumber\end{eqnarray}
The contravariant metric $g^{\nu\mu}$ and skew symmetric
tensor $\omega_{\nu\mu}$
here are equal to
$$
g=U(1)U^T(1),\hspace{1em}
\omega={\Im}\left(\int_{0}^{1}
(U^{-1})^T\frac{\partial}{\partial\tau}\!U^{-1}\, d\!\tau
\right),
 $$
where
$$
U(\tau)=\cosh
\left(\tau
\sqrt{HH^T}
\right)
+H^T\frac{\sinh
\left(\tau
\sqrt{HH^T}
\right)
}
{\sqrt{HH^T}}.
 $$
Local spacetime symmetries
$$
\delta\!g_{\nu\mu}=\varepsilon_{\nu;\mu}
+\varepsilon_{\mu;\nu},\hspace{1em}
\delta\!\omega_{\nu\mu}=
\varepsilon^{\eta}\omega_{\nu\mu;\eta}+
\varepsilon^{\nu;\eta}\omega_{\eta\mu}
+\varepsilon^{\mu;\eta}\omega_{\nu\eta}+d\varsigma_{\nu\mu}
 $$
are a particular case of more general symmetries~(\ref{sym})
with the following choice of parameters:
$$
\xi_z=
:\!(\varepsilon_{\nu}+i\varsigma_{\nu})\partial\!X^{\nu}\!:,
\hspace{1em}
\xi_{\bar{z}}=
:\!(\varepsilon_{\mu}-i\varsigma_{\mu})
\bar{\partial}\!X^{\mu}\!: .
 $$
The contradiagonal components of the energy-momentum tensor can be
shown to be
$$
T_{z\bar{z}}^{\Psi}\approx T_{\bar{z}z}^{\Psi}\approx
-\frac{1}{2}
:\!\left(
R_{\nu\mu}-d\omega_{\nu}\!^{\sigma\rho}d\omega_{\mu\sigma\rho}+
id\omega_{\eta\nu\mu}\!^{;\eta}
\right)
\partial^{\Psi}\!X^{\nu}
\bar{\partial}^{\Psi}\!X^{\mu}\!:.
 $$
Putting here
$\Phi_z=\partial_z^{\Psi}\!:\!\phi\!:$,
$\Phi_{\bar{z}}=\partial_{\bar z}^{\Psi}\!:\!\phi\!:$
we will come to the following equations of motion:
$$
R_{\nu\mu}=
d\omega_{\nu}\!^{\sigma\rho}d\omega_{\mu\sigma\rho}
+2\phi_{;\nu\mu},
\hspace{1em}
d\omega_{\eta\nu\mu}\!^{;\eta}=2\phi^{;\eta}d\omega_{\eta\nu\mu}.
 $$
As a consequence of this equations it can be shown that
$$
2\Box\phi + 4\phi^{;\nu}\phi_{;\nu}=m^2-
\frac{2}{3}d\omega^{\nu\sigma\rho}d\omega_{\nu\sigma\rho}.
 $$
The parameter $m$ here
is the topological constant of the theory
which can  be interpreted as a dilaton mass
It is responsible
for deformation of central charge
$$
c= D+\frac{1}{2}m^2.
 $$
\\[2.5ex]
\noindent
{\Large\bf Acknowledgement}\\[2ex]
\noindent
I am very grateful to Mark Evans for inspiring me to develop the
{\em deformation} approach, for discussions and for many
important suggestions.
It is also a pleasure to thank Toni Weil for significant editing
help.

\end{document}